**Deep Learning of Atomically Resolved Scanning Transmission Electron Microscopy Images: Chemical Identification and Tracking Local Transformations**

Maxim Ziatdinov[1,2,*], Ondrej Dyck[1,2], Artem Maksov[1-3], Xufan Li[1], Xiahan Sang[1], Kai Xiao[1], Raymond R. Unocic[1], Rama Vasudevan[1,2], Stephen Jesse[1,2], and Sergei V. Kalinin[1,2]

[1]*Center for Nanophase Materials Sciences, and* [2]*Institute for Functional Imaging of Materials, Oak Ridge National Laboratory, Oak Ridge TN 37831, USA*

[3]*Bredesen Center for Interdisciplinary Research, University of Tennessee, Knoxville, Tennessee 37996, USA*

* E-mail: ziatdinovma@ornl.gov

**Abstract:**

Recent advances in scanning transmission electron and scanning probe microscopies have opened exciting opportunities in probing the materials structural parameters and various functional properties in real space with angstrom-level precision. This progress has been accompanied by an exponential increase in the size and quality of datasets produced by microscopic and spectroscopic experimental techniques. These developments necessitate adequate methods for extracting relevant physical and chemical information from the large datasets, for which *a priori* information on the structures of various atomic configurations and lattice defects is limited or absent. Here we demonstrate an application of deep neural networks to extract information from atomically resolved images including location of the atomic species and type of defects. We develop a "weakly-supervised" approach that uses information on the coordinates of all atomic species in the image, extracted *via* a deep neural network, to identify a rich variety of defects that are not part of an initial training set. We further apply our approach to interpret complex atomic and defect transformation, including switching between different coordination of silicon dopants in graphene as a function of time, formation of peculiar silicon dimer with mixed 3-fold and 4-fold coordination, and the motion of molecular "rotor". This deep learning based approach resembles logic of a human operator, but can be scaled leading to significant shift in the way of extracting and analyzing information from raw experimental data.

In the last decade, the proliferation of electron microscopy and scanning probe microscopy techniques have generated massive amounts of data on local chemical structure and atomic transformation.[1-3] Since the advent of aberration corrected Scanning Transmission Electron Microscopy (STEM), atomically resolved images of multiple materials classes ranging from multiferroics, semiconductors, and superconductors have become common.[4-8] The further impetus to this field was given by the development of atomically resolved dynamic studies, when the dynamic changes in matter on the atomic level are visualized. These traditionally include the thermal and chemical processes enabled by advanced thermal and environmental holders.[9,10] More recently, progressively more attention is being attracted to the dynamic processes induced by the electron beam irradiation,[11-14] especially promising in the context of e-beam atomic fabrication.[15-17]

Similar advances are achieved in the field of atomically resolved scanning tunneling (STM) and atomic force microscopy (AFM). The recent famous examples include direct imaging of chemical bonds in molecules,[18] visualizing atomic collapse in artificial nuclei on graphene,[19] and inferring mechanisms behind fundamental physical phenomena, such as high-$T_c$ superconductivity, from single atom defect induced scattering patterns.[20] In addition, the STM allows manipulating matter on atomic and molecular levels *via* mechanochemistry[21] or by electron injection from the STM tip.[22]

The characteristic aspect of both STEM and SPM fields is the generation of a large volume of high-veracity experimental data in the form of static or dynamic images. While providing immediate visualization of atomic structures, the interpretation of this data is traditionally limited to qualitative aspects, *e.g.* highlighting features such as presence of structural and topological defects, interfaces, *etc*. In the last decade, a number of approaches emerged based on the quantification of STEM and STM data. In these, atomic coordinates extracted from image analysis are projected on (postulated) mesoscopic order parameter field, providing information on local ferroelectricity, chemical strains, and octahedral tilts. The workflow for such analyses typically comprise the steps of the identification and refinement of atomic positions and subsequent transformation of measured atomic coordinates into the physical quantities. In STEM, this process is enabled by the fact that the ideal image represents a nucleus and the STEM transfer function is usually monotonic and well behaved; in STM and TEM analysis can be more complicated. However, in all cases to date physical models have been postulated by the observer based on the qualitative observations, and quantitative information was obtained within the framework of the imposed model.

In the last several years, advances in machine learning brought significant developments to multiple areas of science and engineering, particularly (qualitative) image recognition.[23,24] The main goal of the machine learning is to generalize from the available training examples to make accurate predictions/classifications on data samples that were not part of the training set. It is currently believed that one of the most promising areas of machine learning is the so-called deep learning (DL), which is the machine learning technique based on (deep) artificial neural networks.[25] The DL models can be used for prediction and classification of 2D and 3D, spatial and temporal data, such as images and videos. Despite that some of the key concepts of artificial neural networks have been researched as early as 1958,[26] it was only several years ago, partially due to overcoming computational limitations, that they gained widespread recognition. In particular, the convolutional neural networks were responsible for major breakthroughs in image and video recognition tasks,[25,27] with seminal breakthroughs such as distinguishing with high accuracy between images of cats and dogs available on the Internet. Since then there was an explosion in applications of DL in various areas of everyday life (*e.g.*, tagging photos on Facebook) as well as in scientific and engineering research (*e.g.* cancer detection, self-driving cars, and satellite imaging). Recently, the DL framework was applied to several problems in theoretical physics such as the search for exotic particles at the Large Hadron Collider[28] and detection of phase transitions in lattice models.[29,30]

Here, we present a "weakly supervised" approach to deep learning of atomically-resolved images, in which by starting with limited *a priori* information about types of defects in a sample, that is, a limited number of available labels in the DL scheme, we use the information on local atomic coordinates to identify a rich variety of defect structures that were not explicitly included in the initial training set. Specifically, by applying Laplacian of Gaussian blob detection technique to the output of the fully convolutional neural network, we were able to make a transition from the classification of image pixels to chemistry-based classification of defects based on the position of chemical species, bond coordination, bond length and bond angle. Using this approach, we identified multiple defect structures associated with silicon atoms implanted into graphene vacancies, as well as observed various chemical and structural transformations of defects on the surface for which we could identify both the location and chemical structure of the defect for each individual step of a "reaction". This approach resembles the approach of a human operator with a background knowledge of the basic materials science concepts but without specific information on the structural and functional peculiarities in each particular material.

**RESULTS & DISCUSSION**

In order to obtain high quality, atomically resolved STEM images for the development and analysis described, we used an aberration corrected Nion UltraSTEM100 operated at 60 kV. Graphene samples were used as a model test case because of their robustness against the 60 kV beam and the ease with which interesting and highly-visible defects may be found or created. Moreover, since these defects are known to evolve/restructure under e-beam influence and each structure is usually only a single atom thick (providing clear delineation of each atom's location), this system offers an ideal test bed for the methods described. In addition, we tested our methods on the STEM images of monolayer $MoSe_2$ doped with tungsten, $Mo_{1-x}W_xSe_2$.

For the analysis of the atomically-resolved experimental data on the model systems we employed the DL neural network model(s) that has an encoder-decoder type of architecture (Fig. 1a).[31] The encoder part consisted of convolutional layers for feature extraction followed by max-pooling layers for reducing the size of processed data. The number of filters in the convolutional layers was doubled after each max-pooling layer. The decoder (or "deconvolutional") part of the network, whose role was to map the encoded low-resolution feature maps to full input-resolution feature maps, consisted of the same filters but in reverse order and *un*-pooling (or up-sampling) layers. The feature maps from the last convolutional layer of the network were passed to a softmax layer for the pixel-wise classification providing us with information on the probability of each pixel being an atom or specific type of defect, depending on the classification scheme used. As we did not employ a fully connected, "dense" layer in our network architecture, these networks are referred to as fully convolutional networks (FCNs).[32]

We train an FCN model using the simulated, theoretical images of atomic lattice and defects and then apply the trained model to the experimental data. This is different from most of the approaches towards training of DL models, which usually utilize some existing database of experimental images with corresponding labels, and hence cannot be used to study materials for which no such databases exist. The theoretical images can be generated using atomic coordinates obtained from *ab-initio* or molecular dynamic simulations of the corresponding atomic structures. In the case of STEM images, the pixel intensities in the generated images are proportional to the atomic number $Z^{1.5-1.8}$ for each atomic column.[33] For the STM images, the intensity of the image is associated with electronic density of states at the Fermi level. In this paper we focus on the analysis of STEM experimental data; however, this approach is applicable to the STM datasets as well.

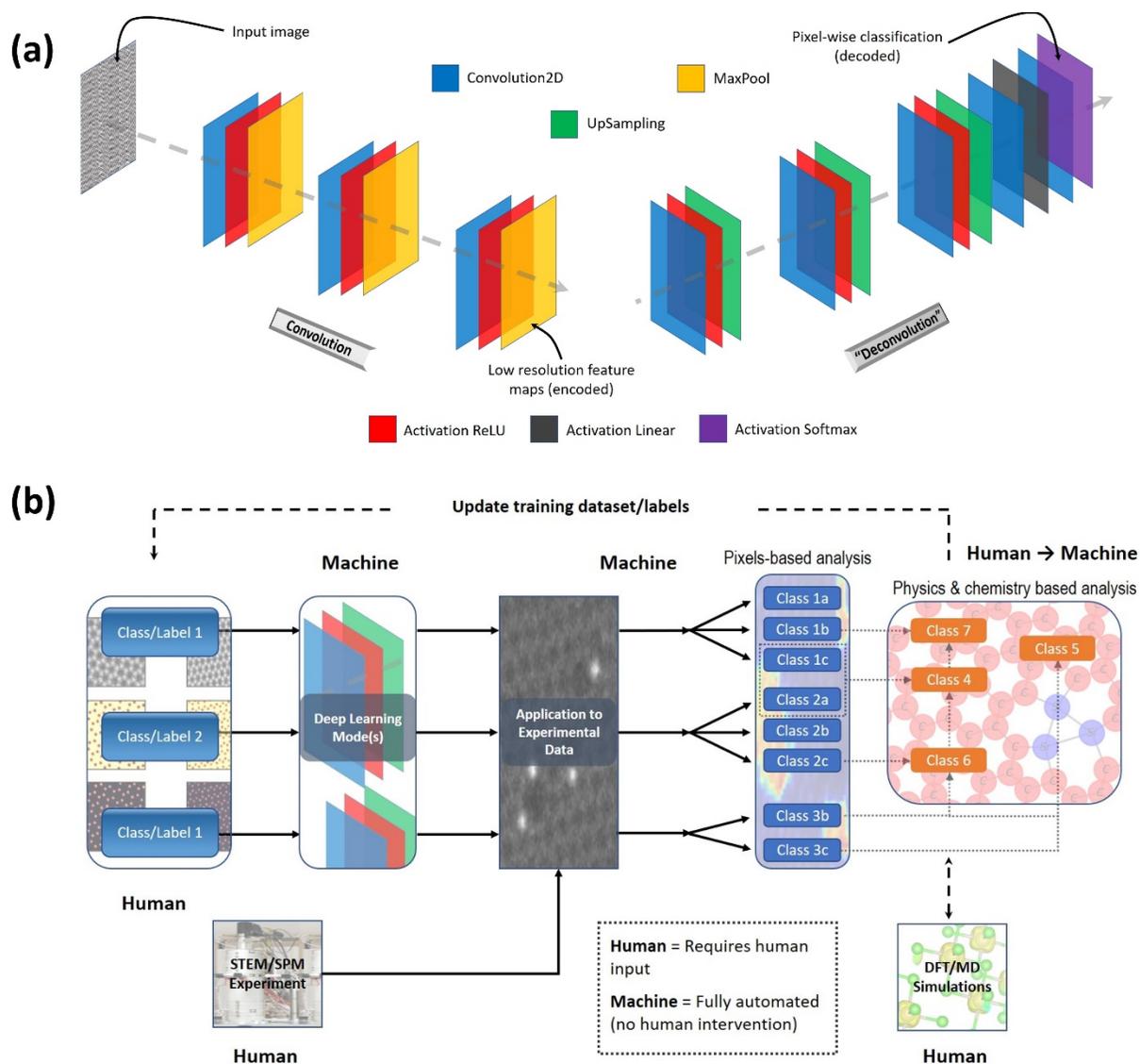

**FIGURE 1. Application of deep learning to a problem of finding atomic species and defects in a crystal lattice.** (a) Schematic architecture of a fully convolutional neural network (FCN) that has an encoder-decoder type of structure (or convolution-"deconvolution" structure). The final softmax layer outputs a pixel-wise classification for atomic species and/or defects. (b) Schematics of a "weakly-supervised" approach towards identifying lattice configurations and defects in the experimental data (see text for details). The parts of the process that require (do not require) human input are denoted as 'Human' ('Machine').

In the following we are going to adopt a learning approach that could be generally referred to as a "weakly-supervised learning". The idea behind this approach is that one starts with a few simple types (classes) of defect structures that enable a sample description in a very general manner (for example, classes associated with lattice and with 1-2 types of defects), trains a classifier, and then applies it to experimental data. In such a case, the output of the DL model will typically show, in addition to the classes that it was trained on, structures that can be an agglomeration of defects belonging to one class (*e.g.* clustering of point defects with various types of "bonding" to the lattice and to each other), those belonging to different classes (*e.g.* clustering of point defects along an extended 1D defect), as well as different subclasses per each class (*e.g.* presence of point defects associated with dopants of different chemical origin). The discovered structures can be then analyzed either qualitatively or quantitatively (*e.g.* through DFT or MD) and used to expand the training set and retrain the DL model. Note that this approach, which is schematically illustrated in Fig. 1b, is very similar to the way human experts explore and discover different types of physical and chemical behaviors in experimental data.

Following the ideas outlined above we start a construction of the training set. First, we note that most of the point (*i.e.* atomic-size) defects in typical STEM and STM datasets can be classified as belonging to two classes, commonly referred to as the depression and protrusion. The former includes missing atoms (STEM and STM), substitutional dopants with lower atomic number $Z$ (STEM), and areas of depleted charge density at the Fermi level (STM), while the latter is commonly associated with adatoms (STEM and STM) and implanted dopants that have either a higher atomic number (STEM) or create additional states at the Fermi level (STM). We therefore prepare the training sets for 3 separate classes, namely, i) lattice without defects, ii) vacancy, and iii) dopant atom (with higher $Z$). The vacancy is created simply by removing a single carbon atom from the lattice. To create a structure with dopant, we remove a carbon atom(s) and place a different atomic element in its position whose intensity is ~2*$I_C$, where $I_C$ is the STEM intensity of carbon atom in the graphene lattice. As our purpose is to "teach" a network only the basic concepts of what vacancies and dopants usually look like in the microscopic images, no relaxation of the lattice structure was performed after removing atom or implanting the dopant atom. We will show later in the paper how such a relatively simple training procedure allows finding and classifying more complex defects in the experimental data even from different materials that were not explicitly included in the training set. To account for a variation in instrumental parameters during image acquisition (*e.g.* different sample orientation, noise and blurring levels) as well as model uncertainties (*e.g.* sample thickness) we perform an augmentation of our training data by applying a set of image transformation operations to our theoretical image(s) (see also Methods and Supplementary Figure 1).

## Pixel-based classification

The application of the trained FCN model(s) to experimental data allows us to identify positions of individual atoms in the lattice as well as location and type of different types of defects. Figure 2 illustrates a procedure for obtaining atomic positions from raw experimental STEM data. First, the raw experimental image is fed into the trained FCN (accuracy on the test set ~ 97 %, see Supplemental Figure 2) that outputs a map of probabilities (*Pr*) for each pixel being an atom (Fig. 2b, e). We then threshold the map at some specific value (usually at *Pr*~0.8-0.9) to produce a binary image with small blobs of a circular shape corresponding to atoms in the lattice. Finally, the coordinates of the atomic centers are calculated *via* a Laplacian of Gaussian (LoG) blob detector technique[34] (Fig. 2c, f).

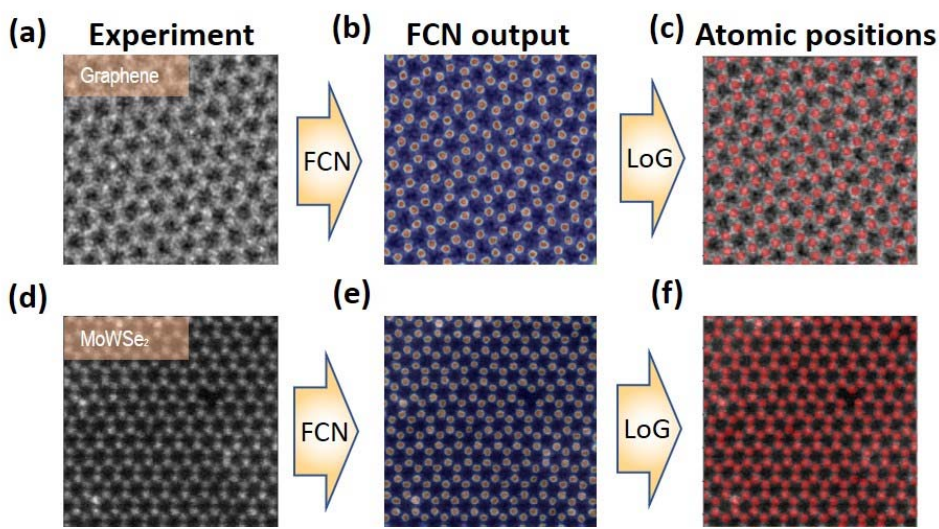

**FIGURE 2. Finding atomic positions from raw experimental images.** (a-f) Illustration of how the atomic positions are obtained from raw experimental data on two different materials, graphene (a-c) and $Mo_{1-x}W_xSe_2$ (d-f), using fully convolutional neural network (FCN) and Laplacian of Gaussian (LoG) blob detection method. The FCN maps and atomic positions are overlaid on the experimental images.

Interestingly, the FCN allows to find atoms in the vertices of hexagons even when atoms do not produce any characteristic local maxima in the image (Fig. 2a-c). This becomes possible because when tasked with finding atomic positions the FCN considers what is known as deep, high-level features in the image such as the shape of an individual hexagon (sides length, angles), as well as shapes of its neighbors. While the FCN was trained on theoretical images of graphene, we demonstrate that it is possible to use it for finding atoms in a different material that has similar but distinct atomic lattice structure. Specifically, we apply our FCN model to locating atoms in the STEM of $Mo_{1-x}W_xSe_2$ that has two inequivalent atomic sites (in terms of their intensities/atomic number) in the unit cell associated with a single Mo atom and with two stacked Se atoms.[8] As we show in Fig. 2d-f the network trained on simple honeycomb structure of graphene can accurately predict locations of atomic centers in the $Mo_{1-x}W_xSe_2$ system.

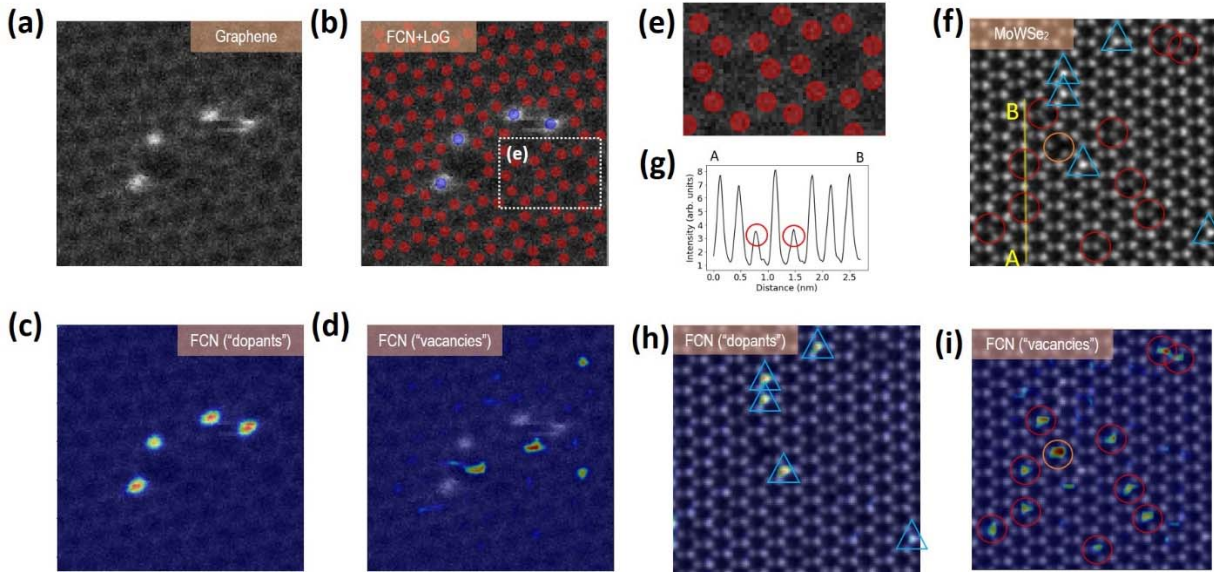

**FIGURE 3. Pixel-wise classification of atomic defects and lattice reconstructions.** (a) Experimental STEM image of graphene with defects. (b) Application of FCN and LoG to extracting locations of atoms and silicon dopants from data in (a). (c, d) FCN output for class "dopants" (c) and class 'vacancy' (d) overlaid on the experimental image. (e) Zoomed area from the box in (b) showing that the DL model trained to find atoms in hexagonal graphene lattice was able to map the 5-7 reconstructions (formation of Stone-Wales defect) in graphene. (f) Experimental STEM image of $Mo_{1-x}W_xSe_2$ ($x \approx 0.18$), where the triangular and circle shapes correspond to the results from FCN outputs for class "dopants" (h) and class "vacancies" (i), respectively, overlaid on the experimental image. The orange circle in (f, i) denote a "full" vacancy with two missing Se atoms, whereas the red circle denotes a "half" vacancy with one missing Se atom. The intensity profile along the AB line in (f) is shown in (g).

We then proceed to the deep learning based analysis of atomic structures in the presence of defects and lattice reconstructions. Figure 3a shows a STEM image of graphene with two types of defects: vacancies (lattice regions with missing carbon atoms) and dopants (Si atoms). The FCN outputs for Si dopants and vacancies are shown in Fig. 3c and 3d, respectively. It is worth noting that while the FCN was trained only on a honeycomb lattice with and without *single* vacancies it was able to i) find vacancies characterized by multiple missing lattice atoms, and ii) identify non-hexagonal, 5-7 reconstructions (Stone-Wales defects[35]) in the graphene lattice near the defect regions (Fig. 3e). Interestingly, in order to find vacancies, a neural network is essentially looking for regions of a surface where the hollow sites are larger than normal (see Fig. 3d). In the Supplemental Material we examine which surface areas a network is usually paying attention to when searching for vacancies and dopants. In general, our analysis of the DL network behavior suggests that the way these networks search for a vacancy, as well as dopants, is similar to the way the human eye would distinguish between regular lattice structure (with top, bridge and hollow sites) and a defect that breaks lattice periodicity.

Similar to our earlier analysis of the clean lattices, we apply the FCN model trained on graphene with defects to find dopants and vacancies in a single layer of a different material, $Mo_{1-x}W_xSe_2$ (Fig. 3f-i). The FCN output for a class "dopants" in $Mo_{1-x}W_xSe_2$ is shown in Fig. 3h. The model is clearly able to detect tungsten dopants which are characterized by a larger intensity in STEM. It is worth noting that the FCN was able to distinguish accurately between the variations in the intensity associated with two different sublattices and that associated with the presence of dopants, even though such specific information was not included in our training dataset. We further analyzed the output of the FCN model for a class "vacancy". Here our model found two types of blobs (denoted by orange and red circles in Fig. 3i) characterized by different intensities (softmax probabilities) and widths and associated with "full" vacancies (two Se atoms missing from a column) and "half" vacancies (one Se atom missing from a column[8]) (Fig. 3g)**.** Importantly, it was not our initial goal to search for the "half" vacancies, nor was the network specifically trained to discover such structures.

## Chemical structure-based classification

In the preceding section, we showed a deep learning based analysis of atomic lattice and defect structures based on classification of image pixels. We now demonstrate that it is possible to use DL networks, in combination with LoG and simple graph representations, to extract relevant structural/chemical parameters such as coordination number, bond lengths and angles, and to classify defects based on their chemical structures.

As a model example, we analyzed a 3-fold and 4-fold coordinated silicon defect implanted in a graphene lattice.[36,37] By applying LoG to thresholded outputs of FCN for class "lattice" and class "defect" we obtained information about positions of atomic centers and locations of defects (Fig. 4a-e). We then extracted coordinates of the lattice atoms and, if applicable, other defect atoms within a specified radius (~3*a*–4*a*, where *a* is a lattice vector) around the dopant atom(s). The extracted structure is represented by a graph $G = (N, E)$. We then applied a number of constraints based on simple chemistry rules to nodes *N* and edges *E* to adjust/modify our graph structure. Specifically, we introduced constraints for a maximum possible number of chemical bonds (edges) for an atom (node) of each type and for a maximum possible length for each type of chemical bond. In addition, we distinguished "edge" lattice atoms bonded to the "foreign" chemical species at the defect sites from the atoms in the bulk. Finally, for composite defects, we ensured that there were no unphysical connections between different parts of the defect, *e.g.* that a node representing a "vacancy" (region where atom(s) is missing) is not connected to nodes representing atomic species. The coordinates of the defect center were calculated as a "center of the mass" of the obtained graph nodes. For images containing more than one defect, the individual defect structures were localized and enumerated by identifying unconnected subgraphs in the graph representation.

This approach allowed obtaining a graph representing a dopant atom(s) and a neighbor lattice (or other dopant) atom(s) to which it is connected directly *via* a chemical bond (Fig. 4c, f) while removing all the "irrelevant" atoms which do not form a direct chemical bond with a dopant atom. Such an approach represents a transition from a pixel-based classification of defects to a classification of defects based on their chemical structure. Indeed, in the case of the 3-fold and 4-fold coordinated Si atoms, the pixel-wise classification based on our initial training set categorized them as the same type of defect. In contrast, our chemistry-based classification scheme allowed identification of an important difference between the two dopants based on the details of their bonding to the carbon lattice. We can also use this approach to extract values of bond lengths and bond angles. For example, we found that a single 3-fold coordinated silicon dopant is characterized by a relatively small distortion in *apparent* Si-C bond lengths, $d_{Si-C} \cong 150$ pm. This can be explained by the presence of out-of-plane distortions of the Si dopant in the 3-fold coordination as predicted by DFT,[37] which is not possible to resolve in the STEM experiment. The angles between projections of Si-C bonds onto a 2D plane are approximately $\alpha_{C-Si-C} \cong 120°$. In the case of 4-fold coordinated Si, the bond lengths and bond angles are $d_{Si-C} \cong 172$ pm and $\alpha_{C-Si-C} \cong 90°$, respectively, which is in general agreement with DFT calculated parameters for planar Si-C structures (*e.g.* 177-179 pm in isoatomic 2D form $Si_{0.5}C_{0.5}$ [38]). The fact that there is such a close agreement between the Si-C bond parameters (length, angle) obtained by our method and those obtained from earlier exhaustive theory-experiment studies for the same defects serves as an additional validation of our model for atomically-resolved STEM experiments.

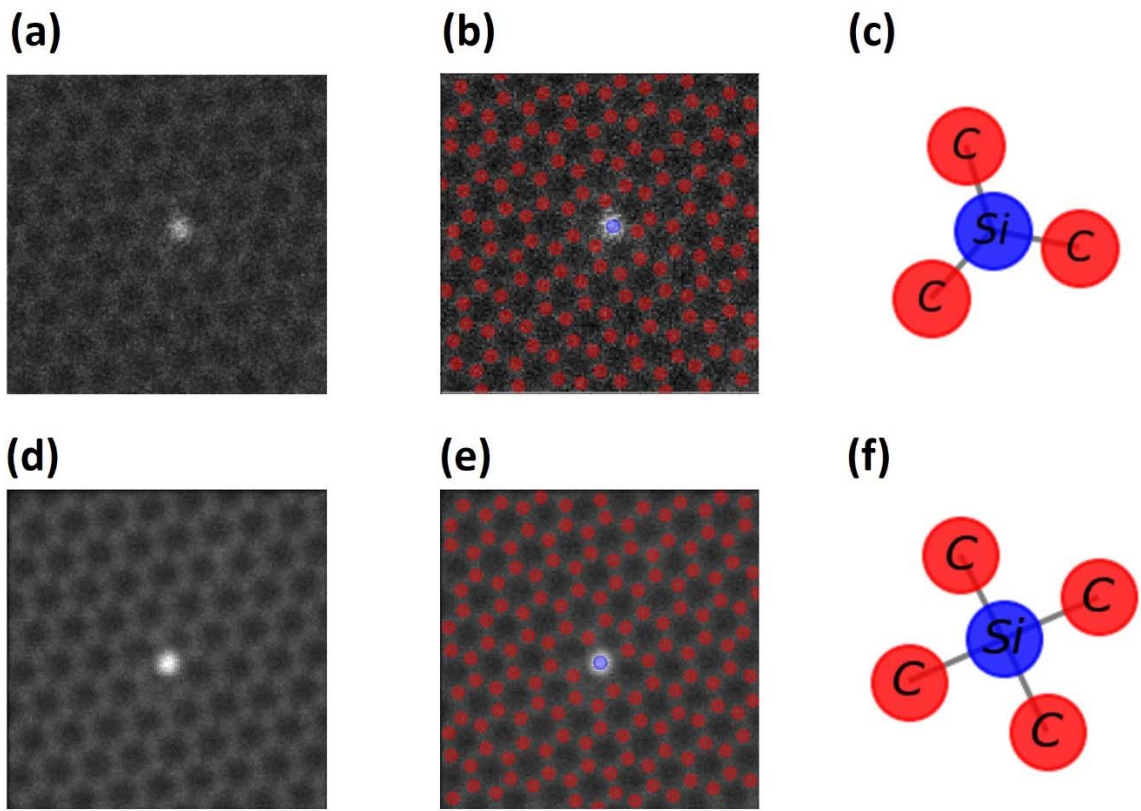

**FIGURE 4. From analysis of pixels in image to analysis of single defect chemistry.** (a) Experimental image of 3-fold coordinated Si defect. (b) Result of applying FCN and LoG to data in (a). (c) Graph representation of single defect structure for 3-fold Si defect (see text for details). (d-f) Same for 4-fold coordinated Si defect.

We further extended the deep learning approach to track the transformation of a Si-lattice defect as a function of time. Specifically, we explored switching from 3-fold to 4-fold coordination and vice versa (Fig. 5) in a frame by frame fashion. The defect class was assigned based on its chemical structure, that is, the number of Si-C bonds, instead of a collection of pixels. This allowed us to obtain both the location of the defect and its type for each frame (Fig. 5b, c). We note that this can be extended to the analysis of more complex surface transformations involving multiple defects in larger image stacks and for larger surface areas.

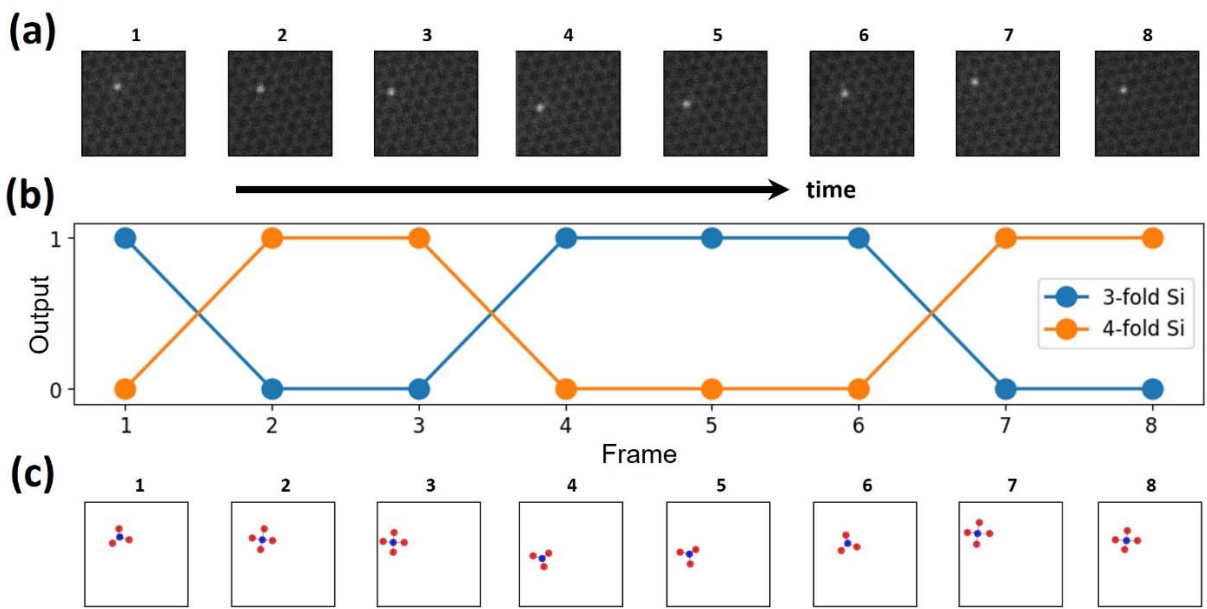

**FIGURE 5. Tracking a reversible switching between different coordinations of Si dopant.** (a) STEM imaging of the same Si dopant over time. (b) Classification of the defect type (3-fold Si vs 4-fold Si) for each frame in (a). (c) Localization of the defect within each frame (the size of the boxes in (c) is identical to that of STEM images in (a)).

We then applied this method to the analysis of complex defect transformations in graphene that involved more than one dopant atom. Figure 6a-d show four consecutive frames of the STEM experiment that starts with an image of a region with a single dopant and ends with an image of the same region but with two dopants that form a dimer-like structure. We first applied FCN and LoG to map the positions of all dopants and lattice atoms. This information was used to construct chemical structures of single defects shown in Fig. 6e-h. The initial configuration was a 4-fold coordinated Si (Fig. 6 a, e). It then underwent a transition to a distorted 3-fold coordinated silicon located next to a carbon vacancy (Fig. 6b, f). The vacancy then moved from right to left inducing a change in the defect orientation that resembled an application of reflection and translation operations to the defect (Fig. 6c, g). The presence of a carbon vacancy resulted in a significant distortion of the $C_3$ symmetry of the defect in frames 2 and 3 (Fig. 6f and 6g) compared to the nearly ideal structure of an isolated defect in Fig. 4a-c. The C-Si-C bond angles were (150°, 115°, 95°) and (165°, 102°, 93°) for defects shown in Fig. 6f and 6g, respectively. We therefore treated the defects in frames 2 and 3 as vacancy-Si-lattice complexes. This represents a situation in which the analysis based on the details of chemical structure of the defects and nearby lattice atoms allowed us to identify a composite type of defect that represents a mixture of two defect classes. Finally, in the fourth frame (Fig. 6d, h) the graphene lattice "captured" another Si atom[16] which resulted in the formation of a peculiar ¾-dimer structure. Here two Si dopants formed a covalent bond with each other and were bonded to two "edge" C atoms ("left" dopant) and three "edge" C atoms ("right" dopant), respectively (Fig. 6h).

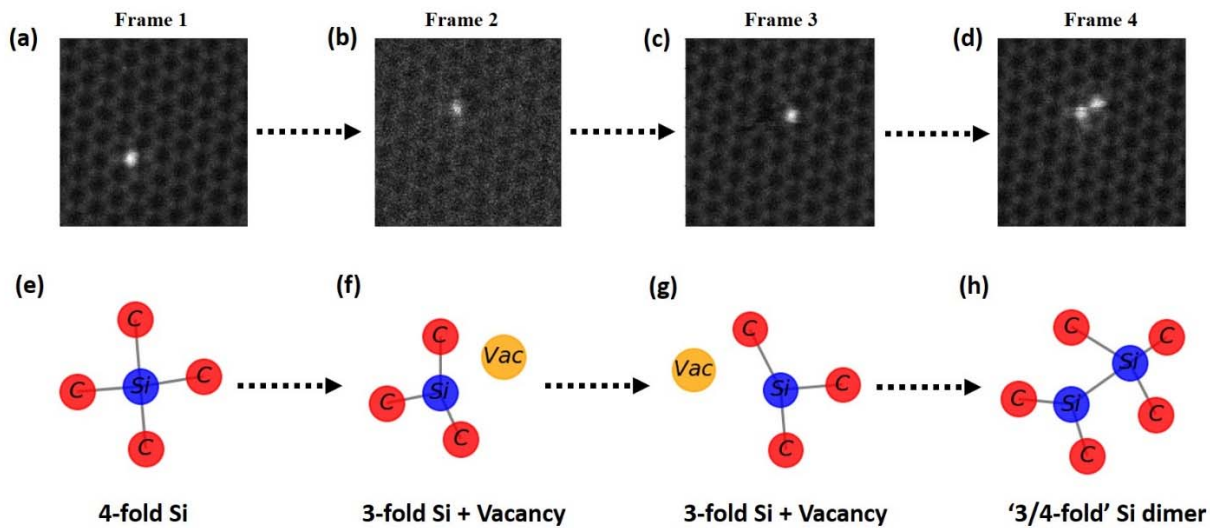

**FIGURE 6. Tracking complex defect transformations on the surface**. (a-d) STEM based tracking of defect transformations on graphene surface. (e-h) Corresponding single defect structures extracted by applying FCN, LoG and graph representations to data in (a-d)

Finally, we extended this framework to the analysis of the defect structures that represent a clustering of Si atoms in the form of a propeller (hereafter, molecular "rotors"). We selected three different frames (Fig. 7a-c) obtained from STEM atomic manipulation of the molecular "rotor". The decoded results representing a single defect chemical structure for each individual case are shown in Fig. 7d-f. The difference between the first frame (Fig. 7a) and the second frame (Fig. 7b) is the rotation of the "rotor" structure by 60°. In the third frame (Fig. 7c), an additional Si atom was placed in the center of the "rotor" structure resulting in a defect structure as shown in Fig. 7f. We found in our analysis that in the 4-Si-rotor structure, the distance between Si atoms in the corners of a triangle increases by ~30% compared to 3-Si-rotor structure, effectively breaking the bonds between corresponding Si atoms. On the other hand, the average length of bonds between the central Si atom and the Si atoms in the corners was comparable to that in carbon lattice, ~ 150 pm. This suggests a scenario in which the Si atom in the center is covalently bonded to three other Si atoms in the corners but is displaced along the $z$-direction which leads to the effective shrinking of the observed bond lengths. We note that as the number of dopant atoms located next to each other within a single defect structure increases, the model trained on a single isolated dopant may return less precise results for the values of bond lengths and bond angles. This, however, can be easily improved by using the identified defects as training samples/labels in the updated model (see Fig.1b).

Having extracted coordinates of all the building blocks of the individual defects we can obtain relevant information on their behavior and evolution over time (*e.g.* spatial trajectories, interaction and 'switching' processes) for larger datasets including STEM 'videos' in a matter of seconds. As an example, we show in Fig. 8 the analysis of the spinning behavior of the molecular "rotor" over the span of 72 STEM frames. Here, we are particularly interested in i) detecting in the automated fashion rotational switching of the "rotor" and ii) analyzing its internal structure such as bond lengths and bond angles for each frame. For the former case, we calculated an angle parameter that describes a relative orientation of three Si atoms with respect to the defect's center of the mass. By taking into account $C_3$ symmetry of the defect, the rotational switching such as the one shown in Fig. 7a,b and 7d,e is described by the change in angle $\Delta\phi = 60°$. Such switching events can be easily seen in Fig. 8a and occur at frames 3-(4)-(5), 8-9, 27-28, 55-56, and 56-57 (Fig. 8b). To analyze the internal structure of the defect we have plotted a deviation of angle in the Si-Si-Si triangle (inset in Fig. 8a) from the case of an equilateral triangle (see also Fig. 7 d,e)). This analysis suggests that while there are some fluctuations in the values of the Si-Si-Si bond angles, typically within ±5°, the overall Si cluster structure remains largely intact. Interestingly, both plots show an anomalous behavior for frame 4. By visualizing this frame for the original experimental dataset, we found that it corresponds to an intermediate step of the rotational switching (*i.e.* the rotation occurred halfway through the image acquisition). We would like to emphasize that this type of analysis is not limited to dopants with higher $Z$ and can be applied to other types of defects, and their motion and transformation, such as movement and rotation of vacancies and vacancy-dopant clusters, once their chemical structures and coordinates are extracted *via* the outlined deep learning based framework.

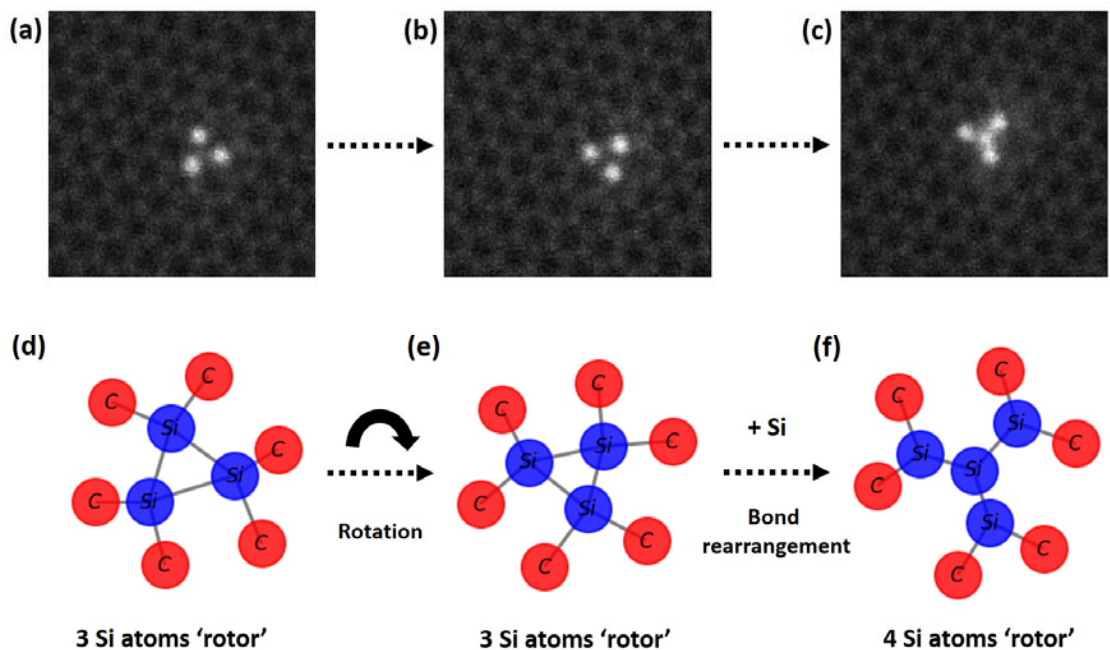

**FIGURE 7. Tracking complex defect transformations on the surface during STEM atomic manipulations.** (a-c). STEM based manipulation of "rotor" structure. (d-f) Corresponding single defect structures extracted, in an automated fashion, after applying FCN, LoG and graph representations to data in (a-c)

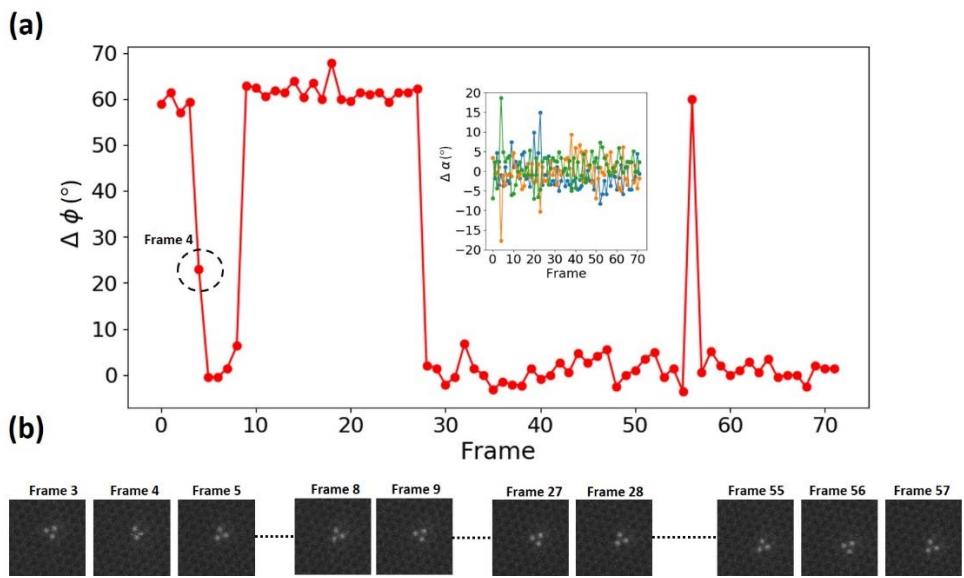

**FIGURE 8. Analysis of "rotor spinning".** (a) Behavior of an angle parameter ($\Delta\phi$) that describes (relative) orientation of the "rotor" in the image; the angle parameter was constructed using the information on coordinates and type of atomic species in the defect, which was extracted by applying FCN and LoG to STEM movie consisting of 72 individual frames (note that count starts from 0). The inset shows deviation from 60° for three angles in the defect's Si-Si-Si triangle. (b) Individual frames of the STEM movie corresponding to the detected switching events.

## CONCLUSIONS

To summarize, we have developed a deep neural network based workflow for identifying positions of atoms in a lattice, type and positions of defect structures, and tracking complex defect transformations from un-processed STEM experimental data. To accommodate limited a priori knowledge about possible defect types (similar to background knowledge of general materials scientist without specific information), we have developed a "weakly supervised" approach, in which we start with just a few simple types (classes) of defects and then use information on local chemical structure to identify a rich variety of defects that were not in the initial training set. In addition, we extracted information relevant to complex defect transformations such as a reversible change between 3-fold and 4-fold coordinated Si dopants in a graphene lattice, formation of a Si dimer consisting of both 3-fold and 4-fold coordinated dopant atoms, and changes in the atomic structure and chemical bonding in the molecular "rotor".

Based on the advances presented in this paper, we envision a development of a "self-driving" microscope for atomically resolved imaging in the not too distant future. In such a microscope, a neural network will be looking through the "eyes" of a microscope on a sample surface and returning information about the position and types of atomic species and defects in real time. Depending on the network's output, the microscope will make necessary actions/adjustments (*e.g.* move the scan frame, or "switch" to the manipulation regime) similar to the way a self-driving car adjusts the steering wheel upon encountering an obstacle on the road.

**Methods:**

**Sample Preparation**

Graphene was grown *via* atmospheric pressure chemical vapor deposition (APCVD) on a Cu foil.[39] The Cu foil was then spin coated with poly(methyl-methacrylate) (PMMA) as a mechanical stabilizer. The Cu foil was dissolved in a bath of ammonium persulfate-deionized (DI) water (0.05 g/ml). The graphene/PMMA film was transferred to hydrogen chloride (HCl) diluted in DI water bath to remove the ammonium persulfate, followed by a DI water rinse. The graphene/PMMA film was placed on a TEM grid and annealed on a hot plate at 150 ºC for ~20 minutes to adhere the grid to the graphene. The PMMA was subsequently dissolved away in an acetone bath, followed by an isopropyl alcohol rinse. Finally the grid was annealed in an Ar-$O_2$ (450 sccm/45 sccm) environment at 500 ºC for 1.5 hours to prevent e-beam induced hydrocarbon deposition in the STEM.[40,41]

2D $Mo_{1-x}W_xSe_2$ monolayers were grown on $SiO_2$/Si substrates at 780 °C by a low pressure CVD method.[42] Poly(methyl methacrylate) (PMMA) was first spun onto the $SiO_2$/Si substrate with monolayer crystals at 3500 rpm for 60 s. The PMMA-coated substrate was then floated on 1M KOH solution, which etched the silica epi-layer, leaving the PMMA film with the monolayer crystals floating on the solution surface. The film was then transferred to deionized water for several times to remove residual KOH. The washed film was scooped onto a Cu TEM grid covered with lacey carbon. The PMMA was then removed with acetone and the samples were then soaked in methanol for 12 h to achieve a clean surface with flakes.

**STEM experiment**

Atomic resolution STEM imaging was performed using a Nion UltraSTEM100 STEM, which is equipped with a spherical aberration ($C_s$) corrector and operated at 60 kV. The high angle annular dark field (HAADF) STEM images were acquired with a convergence angle of 31 mrad and a HAADF detector inner and outer collection angles of 86 mrad and 200 mrad, respectively. The STEM images were introduced to the DL network without any post processing.

**Deep Learning**

All the DL networks employed in our paper have an encoder-decoder type of structure and were implemented using Keras 2.0 (https://keras.io) python deep learning library, with the TensorFlow backend. We found that generally having separate FCN models tailored to find atoms in the lattice and specific defects allows more flexibility when it comes to the analysis of composite defects compared to a single FCN model that is trained to find probabilities for all the classes (lattice + defects) simultaneously. The encoder part of the network for finding lattice atoms/vacancies/dopants typically consisted of 3 convolutional layers with 8/8/2, 16/16/4, and 32/–/– kernels, each of the size 3×3 and stride 1, activated by a rectified linear unit (ReLU) function. Each convolutional layer was followed by a max-pooling layer with a 2 × 2 window and stride 2. The decoder (or "deconvolutional") part of the network consisted of the kernels with the same size and stride plus the un-pooling layers. The last two convolutional layers have the number of kernels equal to the number of classes to be determined, with kernel size 3×3 (pre-last) and 1×1 (last) and stride 1 (note that there is only a linear activation $f(x) = x$ between these two convolutional layers). The feature maps from the final convolutional layer of the network were fed into a softmax classifier for pixel-wise classification, providing us with information on the probability of each pixel being an atom or specific type of defect. The Adam optimizer[43] was used with categorical cross-entropy as the loss function. The atoms in the simulated images were represented as Gaussian "blobs", which a known approximation for representing atomically-resolved STEM images.[44] To generate a large enough training set from theoretical image(s), we applied data augmentation procedure to the original synthetic image(s). Specifically, we used random horizontal/vertical shifts, rotations, zooming-in/out and shear transformations to generate a set containing ~2000 training images for each of the models. Each image was then corrupted with random noise and blurring in the ranges comparable to those typically observed in experiments. Plots of accuracy on the training and validation datasets over training epochs can be found in the Supplemental Material. The input images were resized to 128×128 using pixel area relation. The Laplacian of Gaussian based blob detection applied to the softmax output of the FCN was implemented with scikit-image image processing toolbox (http://scikit-image.org/), and the final graph representations were constructed using NetworkX (http://networkx.github.io). For more details, see the IPython notebook named 'DL STEM' in the Supplementary Material.

**Supporting Information**

The Supporting Information is available free of charge on the ACS Publication website at DOI: *inserted by a publisher*.

Example of training images, model accuracy plots, additional discussion on how the neural networks recognize atomic defects, as well as IPython notebook with a code for constructing and training neural network model, extracting coordinates of atoms/defects, and making graph models.

**Acknowledgements:**

MZ thanks Nouamane Laanait for insightful discussions regarding applications of deep learning in nanoscale imaging. This research was sponsored by the Division of Materials Sciences and Engineering, Office of Science, Basic Energy Sciences, US Department of Energy (MZ, RVK, and SVK). The synthesis of 2D materials (X.L, K.X.) was supported by the U.S. Department of Energy, Office of Science, Basic

Energy Sciences, Materials Sciences and Engineering Division. Research was conducted at the Center for Nanophase Materials Sciences, which is a DOE Office of Science User Facility (XS, RRU), and was also supported by the Laboratory Directed Research and Development Program of Oak Ridge National Laboratory, managed by UT-Battelle, LLC, for the U.S. Department of Energy (OD, SJ). A.M. acknowledges fellowship support from the UT/ORNL Bredesen Center for Interdisciplinary Research and Graduate Education.